\documentclass[aps,prl,twocolumn,superscriptaddress,showpacs,reprint]{revtex4-1}

\usepackage{graphicx}

\begin{document}

\title{The negative thermal expansion mechanism of zirconium tungstate, ZrW$_2$O$_8$}
\author{Leila H. N. Rimmer}

\affiliation{School of Physics \& Astronomy \textit{and} Materials Research Institute, Queen Mary University of London, Mile End Road, London E1 4NS, U.K.}
\affiliation{CrystalMaker Software Limited, Centre for Innovation \& Enterprise, Begbroke Science Park, Woodstock Road, Begbroke OX5 1PF, U.K.}
\affiliation{Department of Earth Sciences, University of Cambridge, Downing Street, Cambridge CB2 3EQ, U.K.}

\author{Martin T. Dove}
\email{martin.dove@qmul.ac.uk}
\affiliation{School of Physics \& Astronomy \textit{and} Materials Research Institute, Queen Mary University of London,  Mile End Road, London E1 4NS, U.K.}
\affiliation{Department of Earth Sciences, University of Cambridge, Downing Street, Cambridge CB2 3EQ, U.K.}

\author{Keith Refson}
\affiliation{Department of Physics, Royal Holloway University of London, Egham Hill, Egham TW20 0EX, U.K.}
\affiliation{Science and Technology Facilities Council, Rutherford Appleton Laboratory, Harwell Science and Innovation Campus, Didcot OX11 0QX, U.K.}

\date{\today}

\begin{abstract}
Negative thermal expansion in ZrW$_2$O$_8$ was investigated using a flexibility analysis of \emph{ab-initio} phonons. It was shown that no previously proposed mechanism adequately describes the atomic-scale origin of negative thermal expansion in this material. Instead it was found that NTE in ZrW$_2$O$_8$ is driven, not by a single mechanism, but by wide bands of phonons that resemble vibrations of near-rigid WO$_4$ units and Zr--O bonds at low frequency, with deformation of O--W--O and O--Zr--O bond angles steadily increasing with increasing NTE phonon frequency. It is asserted that this phenomenon is likely to provide a more accurate explanation for NTE in many complex systems not yet studied.
\end{abstract}


\maketitle

\begin{figure}[hbt]
	\includegraphics[width=0.5\textwidth]{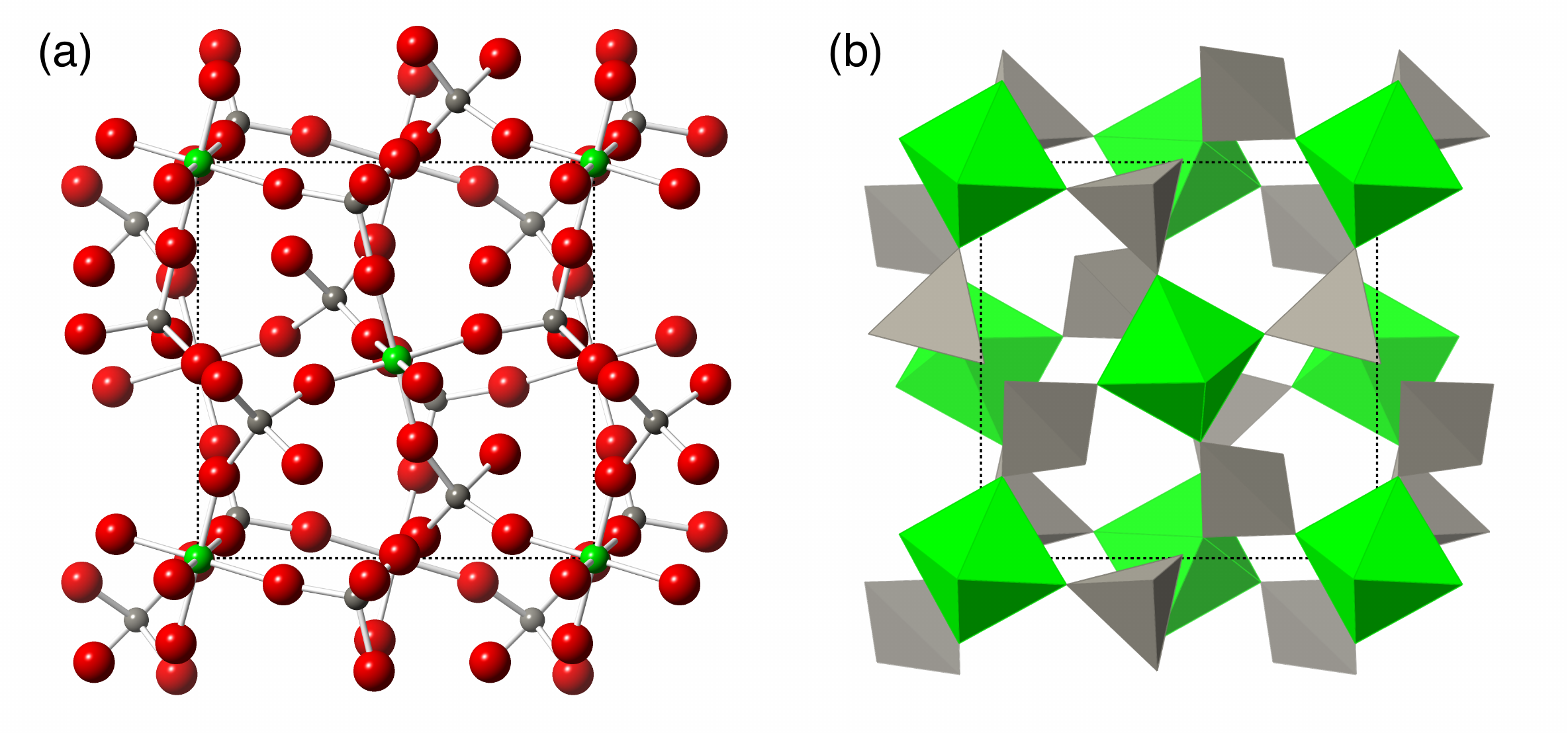}
	\caption{\label{structure}The ZrW$_2$O$_8$ structure. (a) The structure in ball-and-stick format. Green atoms are Zr, grey atoms are W and red atoms are O. (b) The same structure represented as green ZrO$_6$ octahedra and grey WO$_4$ tetrahedra. The space group of this material is $P2_13$.}
\end{figure}

The last two decades have seen a considerable upsurge of interest in materials that have the counter-intuitive and potentially technologically important property of negative thermal expansion (NTE) \cite{Romao_etal_NTEChapterFixedRef,Lind_2012_Mater_NTEReview,Barrera_etal_2005_JPhysCondMat_NTEReview,DoveFangNteReview_RepProgPhys_2016,Takenaka_NTEApplicationsReview,Goodwin_2008_NatNano_NTE}. This interest was prompted by the discovery, in 1996, that NTE in ZrW$_2$O$_8$ (structure shown in Figure \ref{structure}) exists over a wide range of temperatures (0--1050K) \cite{Mary_etal_1996_Science_NTEZrW2O8Experiment,Evans_etal_1996_NTEZrW2O8}. Since then a number of other materials have been shown to exhibit NTE \cite{Romao_etal_2013_NTEChapter,Lind_2012_Mater_NTEReview,Barrera_etal_2005_JPhysCondMat_NTEReview} with research initially focusing on oxides before being extended to other ceramics and hybrid metal-organic materials. There now exists a considerable body of work on the `archetypal' NTE material ZrW$_2$O$_8$, including experiments based on characterization, diffraction and spectroscopy \cite{Ramirez_Kowach_1998_PRL_ZrW2O8LowTCvPeak,Boerio-Goates_etal_2002_JThermAnCal_NTEModesCalorimetry,Yamamura_etal_2002_ZrW2O8Calorimetry,
Ernst_etal_1998_Nature_ZrW2O8NTEPhononDoS,Mittal_etal_2001_PRL_ZrW2O8NTEHighPINS,
Yamamura_etal_2002_ZrW2O8Calorimetry,Ravindran_etal_2000_PRL_ZrW2O8HighPGrunParamThermalProp,
Hancock_etal_2004_ZrW2O8NTERaman,
Tucker_etal_2005_PRL_ZrW2O8NTE,Tucker_etal_2007_JPhysCondMat_ZrW2O8NeutronTotalScattering,
Cao_etal_2002_PRL_ZrW2O8NTEFrustratedModes,Cao_etal_2003_PRB_ZrW2O8NTEXAFS,
Bridges_etak_2014_ZrW2O8XPDFEXAFS,
David_etal_1999_EurophysLett_ZrW2O8NTEMech,Wang_Reeber_2000_ApplPhysLett_ZrW2O8GruneisenCalc,Ravindran_etal_2000_PRL_ZrW2O8HighPGrunParamThermalProp}, together with simulations based on force field and \emph{ab-initio} methods \cite{Mittal_Chaplot_1999_PRB_ZrW2O8LatticeDynamics,
Gava_etal_2012_PRL_ZrW2O8AbInitPhononsGamma,Gupta_etal_2013_PRB_NTEZrW2O8AbInitio,
Pryde_etal_1996_NTEZrW2O8,Pryde_etal_1997_PhaseTrans_ZrW2O8RUMsNTE,
Figueiredo_Perottoni_2007_DFTZrW2O8BondStiffness,
Pryde_etal_1998_JPhysCondMat_NTEZrW2O8OrthorhombicRUMs,
Kojima_etal_2003_ZrW2O8ChargeDensityStudy,
Sanson_2014_ChemMater_ZrW2O8NTENotRums}. However, despite the wealth of studies, NTE in this material is still not understood.

In this letter we offer a significant advance in our understanding of NTE using a new approach we have successfully applied to Zn(CN)$_2$ \cite{Fang_etal_2013_PRB_NTEZnCN2}, Cu$_2$O \cite{Rimmer_etal_2014_PRB_Cu2O}, MOF-5 \cite{Rimmer_etal_2014_PCCP_NTEMOF-5}, and Y$_2$W$_3$O$_{12}$ \cite{Rimmer_Dove_2015_JPCM_NTEY2W3O12}. We model the dynamics of ZrW$_2$O$_8$ in terms of different flexibility models with vibrational modes of these models then being mapped onto the vibrational spectrum of the real material. By observing the degree to which each model is able to recreate the NTE phonons of ZrW$_2$O$_8$, we are able to show that its NTE is caused not by a single mechanism but, rather, by a diverse range of phonons.

Thermal expansion of materials is generally understood in terms of the Gr\"uneisen model, whereby the free energy cost of changing the crystal energy through a change in volume is offset by an increase in entropy through the resultant change in phonon frequencies. 
Normally, the anharmonicity of atomic bonding causes phonon frequencies to increase with a decrease in volume (a positive Gr\"uneisen parameter). However, in the case of NTE materials, we expect to find a significant number of phonons whose frequencies decrease with a decrease in volume (a negative Gr\"uneisen parameter). Given that larger amplitude phonons will have a greater effect on the overall behavior of a crystal, we also expect to find that the most influential NTE phonons would be those that exist at lower frequencies.

\begin{figure}[tb]
	\includegraphics[width=0.5\textwidth]{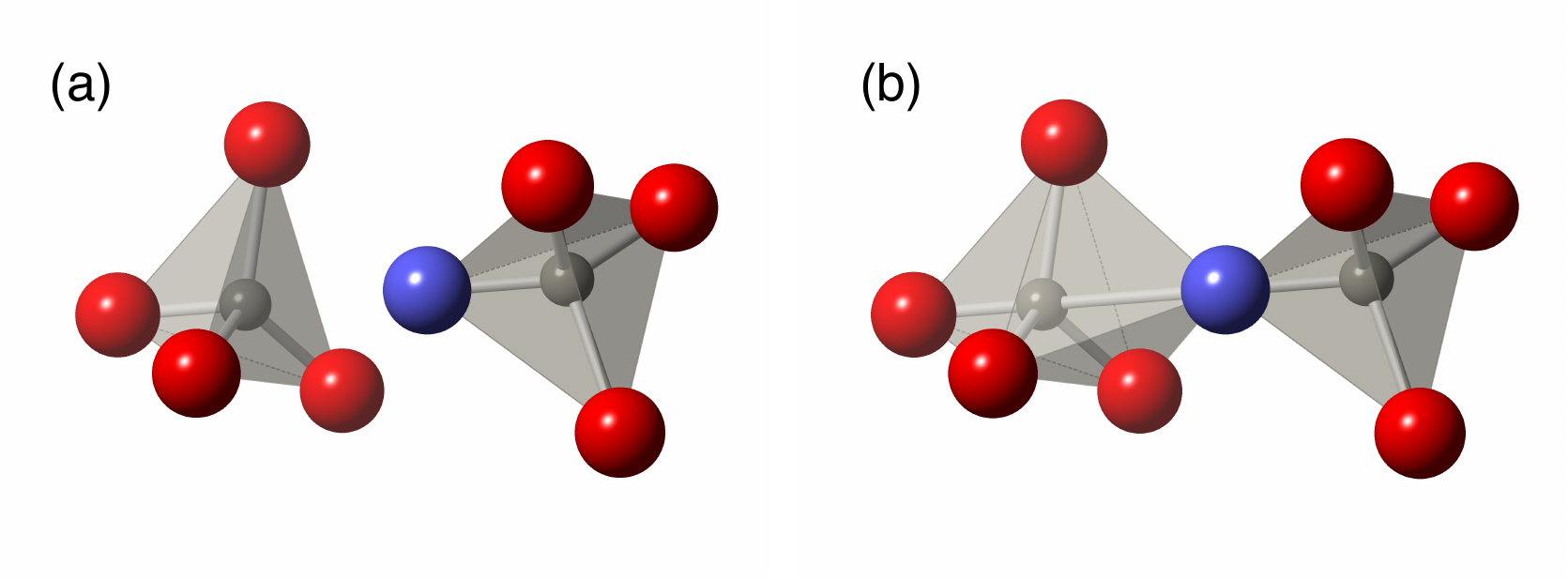}
	\caption{\label{tetpairs}Two interpretations of the interaction between adjacent WO$_4$ tetrahedra in a pair. The blue atom is the under-bonded O that exists between the two W. (a) Both W are tetrahedrally coordinated and there is only weak interaction between the two units (b) The under-bonded O that exists between the two tetrahedra is part of a stronger bond with a second tetrahedron. One W has standard tetrahedral coordination and the other has `4 + 1' coordination.}
\end{figure}

In order to understand the NTE phonons in ZrW$_2$O$_8$, we must identify the types of atomic-scale displacements that characterize them. Experimental and computational evidence to date has led to several different proposals for the NTE mechanism in this material.
These all focus on `tension effects' \cite{Barrera_etal_2005_JPhysCondMat_NTEReview}, which note that Zr--O and W--O bonds  are relatively rigid and, thus, transverse vibrations of an oxygen atom in a Zr--O--W bond might pull the structure inwards and give rise to NTE \cite{Mary_etal_1996_Science_NTEZrW2O8Experiment,Sanson_2014_ChemMater_ZrW2O8NTENotRums,Gupta_etal_2013_PRB_NTEZrW2O8AbInitio}.  The Rigid Unit Mode (RUM) model \cite{Hammonds_etal_1996_AmMineral_QRUMPaper} suggests that, in order to exist at low frequency, the NTE phonons could be tension effect modes that also involve minimal deformation of coordination polyhedra; in this case ZrO$_6$ octahedra and WO$_4$ tetrahedra \cite{Pryde_etal_1996_NTEZrW2O8,Pryde_etal_1997_PhaseTrans_ZrW2O8RUMsNTE,Hancock_etal_2004_ZrW2O8NTERaman,Tucker_etal_2005_PRL_ZrW2O8NTE,Tucker_etal_2007_JPhysCondMat_ZrW2O8NeutronTotalScattering}. An alternative model proposes that the Zr\dots W distance remains unchanged, but significant deformation is allowed in one or more of the ZrO$_6$ and WO$_4$ tetrahedra \cite{Figueiredo_Perottoni_2007_DFTZrW2O8BondStiffness,Cao_etal_2002_PRL_ZrW2O8NTEFrustratedModes,Cao_etal_2003_PRB_ZrW2O8NTEXAFS,Bridges_etak_2014_ZrW2O8XPDFEXAFS}. A further possibility suggests that adjacent WO$_4$ tetrahedra may have some additional interaction between each other \cite{Kojima_etal_2003_ZrW2O8ChargeDensityStudy} which would, in turn, mean that both tetrahedra may move as a single unit for low frequency NTE phonons; this model is illustrated in Figure \ref{tetpairs}. However, thus far, no proposed mechanism has been comprehensively tested against the known NTE phonons in this material---an omission which we now rectify.

For our own investigation, we obtained the phonon spectrum of ZrW$_2$O$_8$ using plane-wave Density Functional Theory calculations in CASTEP \cite{Clark_etal_2005_ZKrist_CASTEP,Segall_etal_2002_JPhysCondMat_CASTEPIntro}. The Perdew-Burke-Ernzehof (PBE) functional \cite{Perdew_etal_1996_PRL_PBEFunctional,Perdew_etal_1997_PRL_PBEFunctionalErrata} was used with optimized norm-conserving pseudo-potentials \cite{RappeOptimizedPP2_PRB_1990,RappeOptimizedPP_PRB_1999}, a plane-wave cutoff energy of 750 eV, and a $3\times3\times3$ Monkhorst-Pack grid \cite{Monkhorst_Pack_1976_PRB_MPGrids} (to carry out reciprocal-space integration of electronic states). The unit cell was relaxed at 0~GPa to a lattice parameter of $a=9.26705$~\AA; atomic coordinates for the relaxed structure can be found in the Supplemental Material. 
The relaxed structure has interatomic distances that are consistent with the experimental structure to within 1--2\%, a result expected for DFT calculations using the PBE functional. 
Phonons were then calculated using Density Functional Perturbation Theory \cite{Refson_etal_2006_PhysRevB_DFPTPaper} with a $3\times3\times3$ Monkhorst-Pack grid of phonon wave vectors and Fourier interpolation of the resulting dynamical matrices (in order to sample the entire Brillouin zone). A further set of phonons were obtained for a second ZrW$_2$O$_8$ cell, relaxed at a fixed off-equilibrium cell parameter of $a=9.2905$~\AA.

\begin{figure}[tb]
	\includegraphics[width=0.5\textwidth]{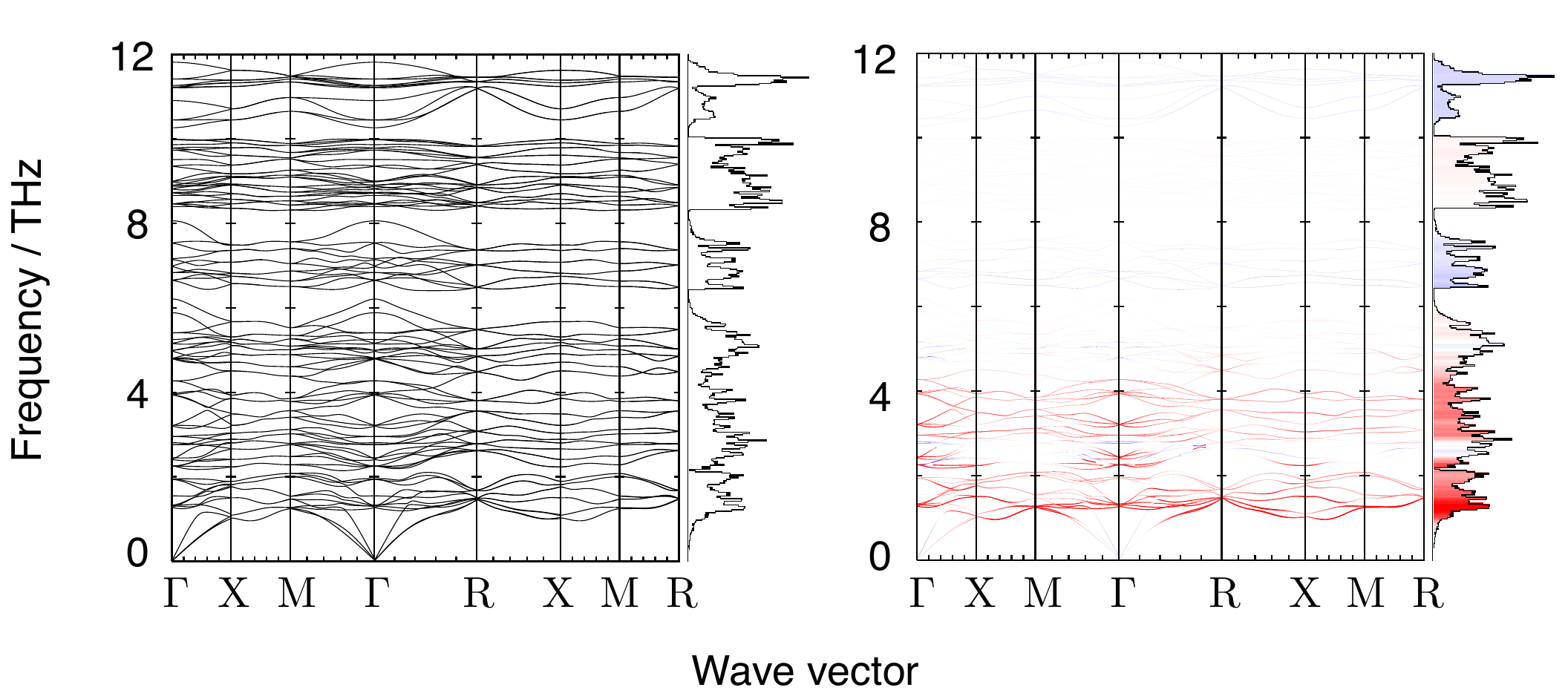}
	\caption{\label{originalphonons}Left: Low energy dispersion curves and densities of states for ZrW$_2$O$_8$. Right: The same data shaded according to the value of $\gamma_{i,\mathbf{k}}$ of each mode at each wave vector. The linear color scale ranges from red ($\gamma_{i,\mathbf{k}}\le-6$) to white ($\gamma_{i,\mathbf{k}}=0$) to blue ($\gamma_{i,\mathbf{k}}\ge6$). Bins that make up the density of states are shaded according to the average $\gamma_{i,\mathbf{k}}$ for each bin.}
\end{figure}

Mode Gr\"uneisen parameters, $\gamma_{i,\mathbf{k}}$, were calculated for each phonon $i$ at each wave vector $\mathbf{k}$ using the approximation
\begin{equation}
\gamma_{i,\mathbf{k}} 
= -\frac{V}{\omega_{i,\mathbf{k}}}\frac{\partial \omega_{i,\mathbf{k}}}{\partial V} 
\approx -\frac{V}{\omega_{i,\mathbf{k}}}\frac{\Delta\omega_{i,\mathbf{k}}}{\Delta V},
\end{equation}
where $V$ is the volume of the equilibrium cell, $\omega_{i,\mathbf{k}}$ is the frequency of a phonon in the equilibrium cell and $\Delta$ indicates the difference in a value calculated for the equilibrium and off-equilibrium cells. Figure \ref{originalphonons}a shows the phonon dispersion curves along high symmetry directions with the phonon density of states for the full Brillouin zone alongside. Figure \ref{originalphonons}b shows the same phonon spectrum shaded according to mode Gr\"uneisen parameter.

NTE phonons span the entire Brillouin zone of this material such that the density of states captures sufficient detail of the phonon spectrum. As a result, for the rest of this letter, phonon spectra are displayed in terms of densities of states with the dispersion curves provided as Supplemental Material. The strongest NTE modes exist around 1.2~THz with NTE character weakening slightly as frequency increases to 2~THz. Bands of weaker NTE modes range from 2--6~THz and from 8--10~THz. There are also some low-frequency, positive thermal expansion (PTE) phonons at wave vectors near $\Gamma$ and X at 2.5~THz and near $\Gamma$ and R at 5~THz. In all, there are 48 modes in the 0--6~THz range which mostly drive NTE (with some driving PTE in the upper frequencies); 12 PTE modes in the 6.5--8~THz range; 28 weak NTE modes in the 8.5--10~THz range; 12 PTE modes in the 10--12~THz range and (not pictured) 32 weak PTE modes above 12~THz.

Flexibility models were constructed as described in \cite{Rimmer_Dove_2015_JPCM_NTEY2W3O12}. Each model is designed such that bonds between atoms are either effectively rigid or completely flexible in order to simulate, in the simplest possible fashion, different types of possible NTE mechanism (e.g.\ a model consisting of rigid Zr--O and W--O bonds with flexible bond angles). 
The ZrO$_6$ unit was modelled in one of two ways: either as a rigid ZrO$_6$ octahedron, or else as rigid Zr--O bonds with flexible O--Zr--O angles. The WO$_4$ unit was modelled in a similar fashion: either as a rigid WO$_4$ tetrahedron, or else as rigid W--O bonds with flexible O--W--O angles. Models to take account of possible bonding between adjacent WO$_4$ units, as per Figure \ref{tetpairs}, were also considered. Bonding between adjacent rigid WO$_4$ tetrahedra was modelled in one of three ways: no bonding, as a rigid bond between adjacent rigid WO$_4$ with a flexible W--O--W angle, or else as a rigid bond between adjacent rigid WO$_4$ with a rigid W--O--W angle.

Models were implemented in the program GULP \cite{Gale_Rohl_2003_MolSim_GULP} using harmonic potential energy functions with large force constants to define rigid bonds and zero force constant to define flexible bonds. No additional functions, such as Coulomb interactions, were used. Phonons were calculated for each model at the same wave vectors as used in the \emph{ab-initio} calculations. Given the manner in which the flexibility models were constructed, any phonon that does not require deformation of any of the model's designated rigid units has zero frequency. The frequencies $\omega$ and eigenvectors $\mathbf{e}$ of the model phonons $j$ were then mapped on to the \emph{ab-initio} phonons $i$ at each wave vector using the equation
\begin{equation}\label{DRAWmatch}
m_{i,\mathbf{k}} = \Omega^2\sum_j\frac{\mathbf{e}_{i,\mathbf{k}}\cdot\mathbf{e}_{j,\mathbf{k}}}{\Omega^2+\omega^2_{j,\mathbf{k}}}
\end{equation}
where $\Omega$ is a constant. A $m_{i,\mathbf{k}}$ value of 1 indicates that the flexibility model is able to perfectly recreate the \emph{ab-initio} phonon in question while a $m_{i,\mathbf{k}}$ value of 0 indicates that there is no relation between the two. Further information on the use of this type of mapping can be found in \cite{Rimmer_Dove_2015_JPCM_NTEY2W3O12}. 

It was found that four models were too constrained and could only generate $m_{i,\mathbf{k}}$ values of zero. These were the three models involving rigid WO$_4$ tetrahedra with a rigid Zr\dots W rod, as well as the model involving rigid ZrO$_6$ octahedra, rigid WO$_4$ tetrahedra, a rigid bond between adjacent WO$_4$, and a rigid W--O--W bond angle. This implies that no part of the lattice dynamics of ZrW$_2$O$_8$ is consistent with these constraints and, thus, these models cannot be considered to be associated with NTE.

\begin{figure}[tb]
	\includegraphics[width=0.5\textwidth]{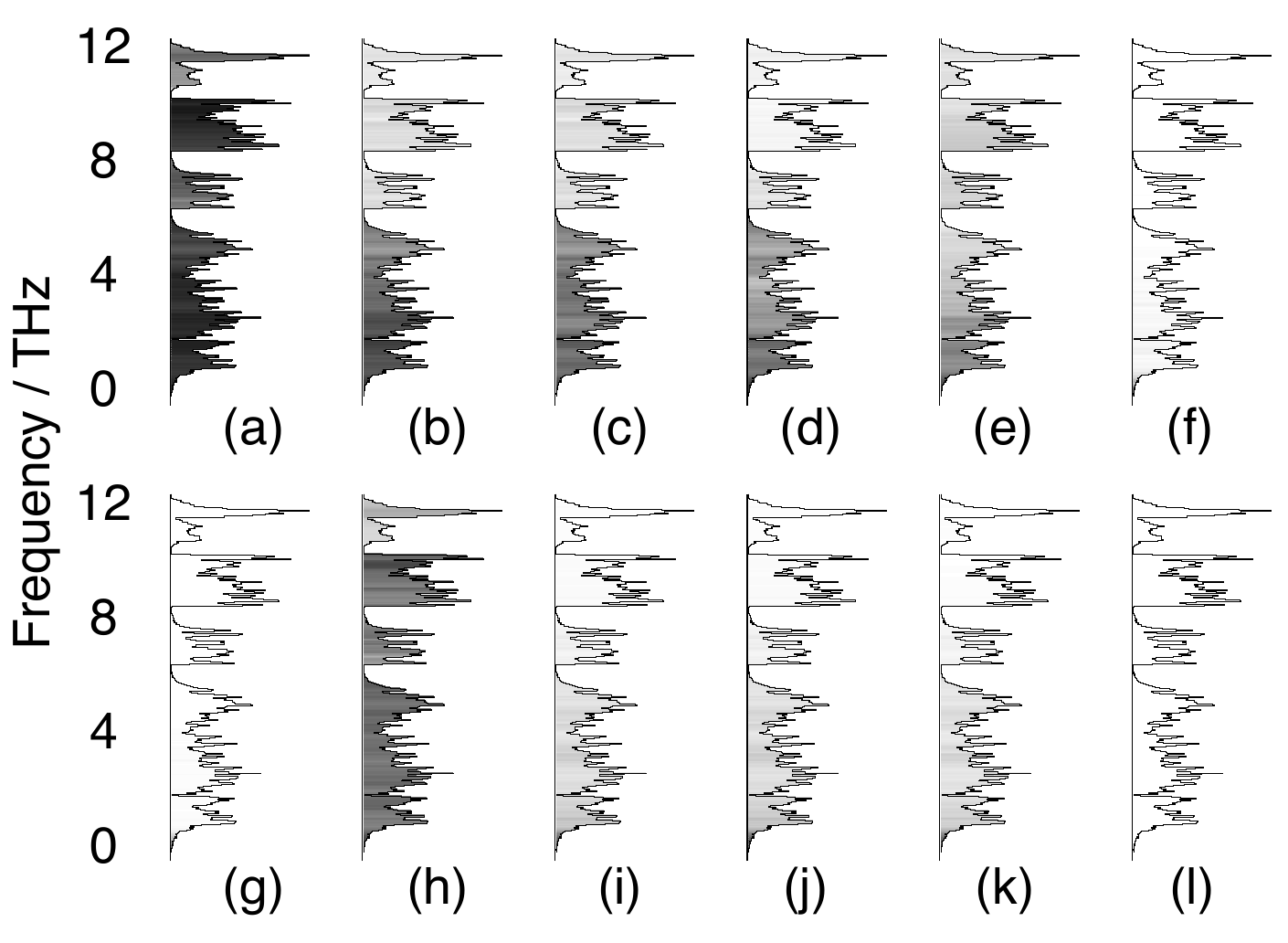}
	\caption{\label{dosplots}Flexibility analysis of ZrW$_2$O$_8$ phonon densities of states. Data is shaded according to the average $m_{i,\mathbf{k}}$ value for each bin. The linear scale ranges from white ($m_{i,\mathbf{k}}=0$) to black ($m_{i,\mathbf{k}}=1$). Models tested are as follows: (a) Rigid Zr--O, W--O (the tension effect). (b) Rigid Zr--O, WO$_4$. (c) Rigid Zr--O, WO$_4$; bonded but flexible W--O--W. (d) Rigid Zr--O, WO$_4$; bonded and rigid W--O--W. (e) Rigid ZrO$_6$, W--O. (f) Rigid ZrO$_6$, WO$_4$ (the RUM model). (g) Rigid ZrO$_6$, WO$_4$; bonded but flexible W--O--W. (h) Rigid Zr--O, W--O, Zr\dots W. (i) Rigid Zr--O, WO$_4$, Zr\dots W. (j) Rigid Zr--O, WO$_4$, Zr\dots W; bonded but flexible W--O--W. (k) Rigid Zr--O, WO$_4$, Zr\dots W; bonded and rigid W--O--W. (l) Rigid ZrO$_6$, W--O, Zr\dots W.}
\end{figure}

Figure \ref{dosplots} shows the \emph{ab-initio} phonon density of states with bins shaded according to the average $m_{i,\mathbf{k}}$ for each model. Unlike simpler systems such as Zn(CN)$_2$ or Cu$_2$O, flexibility model eigenvectors do not neatly map onto a small number of specific modes in the \emph{ab-initio} system. Instead, a high degree of eigenvector mixing is present wherein a given type of flexibility is spread over multiple \emph{ab-initio} phonons, a phenomenon made possible by the low symmetry of the structure. Eigenvector mixing is particularly evident for the rigid ZrO$_6$ and WO$_4$ model (i.e.\ the RUM model) mapped in Figure \ref{dosplots}f where the RUMs are spread so widely across the phonon spectrum that no mode has a $m_{i,\mathbf{k}}$ anywhere approaching 1.

These results show that the 48 phonons in the 0--6~THz range correspond to motion of near-rigid WO$_4$ tetrahedra and Zr--O rods. O--W--O and O--Zr--O bond bending are minimized (although the O--Zr--O angle distorts more) at the lowest frequencies. As frequency increases, so too does the degree of both types of bond bending, with O--Zr--O bond bending increasing steadily, whilst O--W--O bond bending increases slowly before experiencing a jump at 6~THz. This implies that the O--W--O bond angle is stiffer than the O--Zr--O, even though both distort. Neither the RUM model nor the rigid Zr\dots W model correlates well with any of the NTE modes.

Some NTE modes around 4~THz involve additional bending of the W--O--W angle between two WO$_4$ units in a pair. However, since there is no effect on Gr\"uneisen parameter, this additional angle bending appears to make no extra contribution to thermal expansion behavior. Meanwhile the small number of PTE modes in the 0--6~THz range appear to correlate with changes in the distance between two WO$_4$ units in a pair at 2.5~THz and to a small amount of bond stretching within the coordination polyhedra at 5~THz. Upon further inspection of the dispersion curves in the Supplemental Material both sets of modes (and thus the PTE that they drive) are highly wave vector dependent, appearing around $\Gamma$ and X, and at $\Gamma$ and R points respectively. The fact that the occasional separation of adjacent WO$_4$ units does not significantly affect mode frequency suggests that there is no strong bonding between them, but that they do move in tandem more often than not (most likely as a consequence of constraints imposed by the wider ZrW$_2$O$_8$ framework).

The 12 PTE phonons in the 6--8~THz range involve further O--Zr--O and O--W--O bond angle bending as well as a significant amount of Zr--O and W--O bond stretching. This latter effect is a well-known PTE mechanism and is what ultimately gives these phonons their PTE character. The 28 weak NTE modes in the 8--10~THz range involve negligible bond stretching but increased bending of both O--Zr--O and O--W--O bond angles. This band of modes most closely resembles the traditional tension effect, however they exist at high frequency and thus only weakly drive NTE due to the large amount of energetically costly bond bending they incur. The 12 phonons in the 10--12~THz range comprise another band of weak PTE modes that involve similar distortions to those seen in the 6--8~THz range but with a much greater degree of bond stretching that gives these modes both their high frequency and their PTE behavior. PTE modes above 12~THz consist of the remaining bond stretches.

Thus, NTE in ZrW$_2$O$_8$ cannot be described by a single mechanism (such as a tension effect or a model derived therefrom). Instead, it is driven by a broad spectrum of phonons extending over a wide range of energies. The strongest of these exist around 1.2~THz and involve minimal deformation of the WO$_4$ and ZrO$_6$ polyhedral units, with the O--Zr--O bond deforming more than the relatively rigid O--W--O bond. As frequency increases, the level of O--Zr--O and O--W--O bond bending gradually increases to eventually become a mode resembling transverse vibrations of rigid W--O and Zr--O bonds. 

The behavior of ZrW$_2$O$_8$ is reminiscent of that found in Y$_2$W$_3$O$_{12}$ \cite{Rimmer_Dove_2015_JPCM_NTEY2W3O12} and, taken together with recent work on other NTE materials \cite{Rimmer_Dove_2015_JPCM_NTEY2W3O12}, strongly suggests that the widely accepted explanation of NTE in terms of simple mechanisms is not sufficient to explain this phenomenon across all framework materials.

In the simplest of cases such mechanisms can provide an elegant description of observed behavior. For instance, the specific modes shown to drive NTE in ScF$_3$ \cite{Li_etal_2011_ScF3NTE}, Zn(CN)$_2$ \cite{Fang_etal_2013_PRB_NTEZnCN2} and Cu$_2$O \cite{Rimmer_etal_2014_PRB_Cu2O} all look like RUMs. This seems logical given that deformation of a coordination polyhedron is an energetically costly process and a tension effect which minimises this deformation will necessarily exist at low frequency. However with added complexity, such as a large number of phonon branches in a low symmetry structure, we find that (despite being supported by the structure) simple mechanisms based on tension effects or RUMs are not sufficient to explain a material's NTE behavior. Mode mixing is always present in the vibrational spectra of real materials but, in less complex high symmetry structures, these `deeper' developments are masked and thus simpler mechanisms can be sufficient to explain NTE. It is only via an in-depth study of NTE framework materials with greater structural complexity that the more complete picture can be drawn out.

Since this is a behavior that can be expected to occur in other structurally complex NTE materials,  the authors offer the identification of this more complete explanation of NTE in ZrW$_2$O$_8$ as the key contribution of this letter to a general understanding of NTE across all framework materials.

\begin{acknowledgements}
The authors are grateful to Toby Perring for initiating the DFT calculations. LHNR has received support from NERC (Grant No. NE/I528277/1), Innovate UK (Grant No. KTP009358), and CrystalMaker Software Ltd. KR has received support from EPSRC (Grant No. EP/F036809/1) and the STFC Scientific Computing Department for computing time on HECToR and SCARF facilities, respectively. 
\end{acknowledgements}

\end{document}